\documentclass[amsmath,amssymb,aps,prapplied,nonbibnotes,longbibliography,reprint, superscriptaddress]{revtex4-2}
\usepackage{bm}
\usepackage[breaklinks=true,colorlinks=true,linkcolor=blue,urlcolor=blue,citecolor=blue]{hyperref}
\usepackage{soul,color}
\usepackage{graphicx}
\newcommand{\units}[1]{\ensuremath{\,\mathrm{#1}}}

\begin{document}

\title{Vortex matching at 6\,T in YBa$_2$Cu$_3$O$_{7-\delta}$ thin films by imprinting a 20\,nm-periodic pinning array with a focused helium ion beam}

\author{Max~Karrer}
  \affiliation{Physikalisches Institut, Center for Quantum Science (CQ) and LISA$^+$, University of T\"ubingen, Auf der Morgenstelle 14, 72076, T\"ubingen, Germany}
\author{Bernd~Aichner}
  \affiliation{Faculty of Physics, University of Vienna, Boltzmanngasse 5, 1090 Vienna, Austria}
\author{Katja~Wurster}
 \affiliation{Physikalisches Institut, Center for Quantum Science (CQ) and LISA$^+$, University of T\"ubingen, Auf der Morgenstelle 14, 72076, T\"ubingen, Germany}
\author{C\'esar Mag\'en}
  \affiliation{Instituto de Nanociencia y Materiales de Aragón (INMA), CSIC-Universidad de Zaragoza, 50009 Zaragoza, Spain}
\author{Christoph Schmid}
   \affiliation{Physikalisches Institut, Center for Quantum Science (CQ) and LISA$^+$, University of T\"ubingen, Auf der Morgenstelle 14, 72076, T\"ubingen, Germany}
\author{Robin Hutt}
   \affiliation{Physikalisches Institut, Center for Quantum Science (CQ) and LISA$^+$, University of T\"ubingen, Auf der Morgenstelle 14, 72076, T\"ubingen, Germany}
\author{Barbora~Budinsk\'a}
  \affiliation{Faculty of Physics, University of Vienna, Boltzmanngasse 5, 1090 Vienna, Austria}
  \affiliation{Vienna Doctoral School in Physics, University of Vienna, Boltzmanngasse 5, 1090 Vienna, Austria}
\author{Oleksandr~V.~Dobrovolskiy}
  \affiliation{Faculty of Physics, University of Vienna, Boltzmanngasse 5, 1090 Vienna, Austria}
  \affiliation{Cryogenic Quantum Electronics, Institut für Elektrische Messtechnik und Grundlagen der Elektrotechnik, Technische Universit\"at Braunschweig, Hans-Sommer-Str. 66, 38106 Braunschweig, Germany and Laboratory for Emerging Nanometrology, Technische Universit\"at Braunschweig, Langer Kamp 6a/b, 38106 Braunschweig, Germany}
\author{Reinhold~Kleiner}
  \affiliation{Physikalisches Institut, Center for Quantum Science (CQ) and LISA$^+$, University of T\"ubingen, Auf der Morgenstelle 14, 72076, T\"ubingen, Germany}
\author{Wolfgang~Lang}
 \email{wolfgang.lang@univie.ac.at}
 \affiliation{Faculty of Physics, University of Vienna, Boltzmanngasse 5, 1090 Vienna, Austria}
 \author{Edward~Goldobin}
\author{Dieter~Koelle}
  \email{koelle@uni-tuebingen.de}
  \affiliation{Physikalisches Institut, Center for Quantum Science (CQ) and LISA$^+$, University of T\"ubingen, Auf der Morgenstelle 14, 72076, T\"ubingen, Germany}

{\onecolumngrid
This is an author-created version of an article published in Phys.~Rev.~Applied~{\bf 22},~014043~(2024). The Version of Record is available online free of charge under the \href{https://creativecommons.org/licenses/by/4.0/}{CC BY 4.0} license at \href{https://journals.aps.org/prapplied/pdf/10.1103/PhysRevApplied.22.014043}{https://journals.aps.org/prapplied/pdf/10.1103/PhysRevApplied.22.014043}. \hrule}


\begin{abstract}
Controlled engineering of vortex pinning sites in copper-oxide superconductors is a critical issue in manufacturing devices based on magnetic flux quanta. To address this, we employed a focused He-ion beam (He-FIB) to irradiate thin YBa$_2$Cu$_3$O$_{7-\delta}$ films and create ultradense hexagonal arrays of defects with lattice spacings as small as 20\,nm. Critical current and magnetoresistance measurements demonstrate efficient pinning by an unprecedentedly high matching field of 6\,T visible in a huge temperature range from the critical temperature $T_c$ down to 2\,K. These results show that He-FIB irradiation provides excellent opportunities for the development and application of superconducting fluxonic devices based on Abrikosov vortices. In particular, our findings suggest that such devices can operate at temperatures far below $T_c$, where superconductivity is robust.
\end{abstract}

\maketitle

\section{Introduction}

Superconductivity is a macroscopic quantum phenomenon characterized by the long-range coherence of charge carriers that make up the superconducting condensate. Most technically viable superconductors, including copper-oxide superconductors (HTS) with high critical temperature $T_c$, belong to the class of type-II superconductors. These materials allow the magnetic field to penetrate in the form of Abrikosov vortices of supercurrent, which carry the quantized magnetic flux $\Phi_0 = h/(2e)$, where $h$ is Planck's constant and $e$ is the elementary charge.

The properties and the interaction of Abrikosov vortices are controlled by two characteristic lengths: the Ginzburg-Landau coherence length $\xi$ and the London penetration depth $\lambda$ \cite{BLAT94R}. In the vortex center, the superconducting order parameter is suppressed to zero on the length scale $\xi$, thus creating a normal-conducting vortex core of approximately $2\xi$ diameter. As the formation of this normal-conducting core (i.e., the suppression of superconductivity) requires energy, the vortex will be preferentially pinned to a nonsuperconducting defect.

The magnetic field and the supercurrents circulating around the vortex core decay exponentially on a characteristic scale given by the London penetration depth $\lambda>\xi$, which determines the magnetic vortex-vortex interaction range. For thin superconducting films with thickness $t_z\lesssim\lambda$, the so-called Pearl length \cite{PEAR64} $\Lambda = 2\lambda^2/t_z$ replaces $\lambda$.

Importantly, Ginzburg-Landau theory predicts the temperature dependence $\xi(T)=\xi(0) (1-T/T_c)^{-1/2}$, which implies that $\xi(T)\to\infty$ at temperature $T \to T_c$. The same temperature dependence is predicted for $\lambda(T)$ \cite{BLAT94R}. Consequently, at temperatures near $T_c$, the size of the vortex cores and their interaction range rises significantly.
For the anisotropic compound YBa$_{2}$Cu$_{3}$O$_{7-\delta}$ (YBCO) used here as a prototypical HTS, the in-plane parameters are $\xi_{ab}(0) \approx 1.2\units{nm}$ and $\lambda_{ab}(0) \approx 250\units{nm}$ (for epitaxial thin films) \cite{SEKI95,ROHN18,Martinez-Perez17}.

A single Abrikosov vortex can store a classical data bit \cite{GOLO15} and can be manipulated by several methods \cite{KERE23}. Vortices and interacting vortex ensembles have been suggested to build superconducting devices like vortex ratchets \cite{WAMB99}, cellular automata \cite{HAST03,MILO07}, superconducting diodes, vortex-based memory cells \cite{GOLO23} and artificial vortex ice \cite{LIBA09}. Although some of these concepts have been realized in experiments \cite{VILL03,LYU21,TRAS14}, it has not yet been possible to exploit the fundamentally possible degree of complexity and miniaturization. Hence, previous experiments with artificial vortex pinning sites were restricted by the employed etching techniques and had to be operated at temperatures $T \to T_c$ where the interaction length $\Lambda(T)$ is longer than the spacing $a$ between the pins \cite{MOSH11M,CAST97,AVCI10}. Innovative concepts, such as growing Nb films on nanopore templates \cite{VINC06}, have been employed to mitigate these issues.

However, one aims to create devices that work at temperatures below about $T_c/2$ where superconductivity is robust and thermodynamic fluctuations do not reduce the collective vortex effects. Accordingly, the goal is to design and fabricate arrangements of many pinning sites in YBCO with diameters $d \gtrsim 2\xi_{ab}(T_c/2)$ and spacings $a$ such that $d < a \ll \Lambda(T_c/2)$.

Thin films of YBCO are usually patterned by etching away material, which leads to a landscape with varying thickness. This process is extremely delicate and becomes even more challenging for nanoscale patterning due to the potential loss of oxygen through the open side faces of the remaining material. However, the loss of oxygen ultimately results in the degradation of the superconducting properties.

To overcome this issue, a masked light-ion beam from an implanter has been employed to fabricate regions with suppressed superconductivity \cite{KAHL98,KATZ00,KANG02a,SWIE12,LANG06a,HAAG14,ZECH17a,ZECH18,ZECH18a}. The impacting ions create mainly Frenkel defects of the oxygen atoms in the YBCO crystal \cite{GRAY22}, while maintaining the crystallographic framework and the surface of the film intact. The binding energies are typically in the range of 1--3\,eV \cite{ROTH89,TOLP96} for O atoms in the CuO chains and about $8\units{eV}$ \cite{TOLP96} for the O atoms in the CuO$_2$ planes of YBCO. Point defects caused by ion impact break the superconducting carrier pairs and suppress $T_c$. These defects also scatter charge carriers and increase the normal-state resistivity. He$^+$ ions with moderate energy are ideal \cite{LANG09} for creating enough point defects to completely suppress superconductivity at a practically available fluence of $\phi < 10^{16} \,\units{ions/cm}^{2}$ while still penetrating a YBCO film with a thickness of less than about 100\,nm \cite{LANG10R}. However, columns of nonsuperconducting material of diameters \cite{TRAS14} $d$ and spacings \cite{YANG22} $a$  achieved by masked ion irradiation of YBCO are presently limited to $d \sim 70\units{nm}$ and $a \sim 100\units{nm}$.    
 
In this work, we utilize the focused He-ion beam (He-FIB) of a helium ion microscope (HIM) to achieve a spatial resolution well below 70\,nm. We exemplify the advantages of the HIM's unique properties \cite{HLAW16M} by showcasing vortex-matching effects in a set of hexagonal pinning arrays. The matching effects show up in measurements of the critical current $I_c$ and resistance $R$ versus magnetic field $B$, and they are remarkably robust, extending to $B$ as high as 6\,T, and persisting at $T$ from $T_c$ down to 2 K. Our results represent a significant advancement in the fabrication of vortex pinning landscapes in thin YBCO films.

\section{Experimental methods}

\subsection{Sample fabrication}

\begin{figure*}[!htb]
  \begin{center}
     \includegraphics[width=\textwidth]{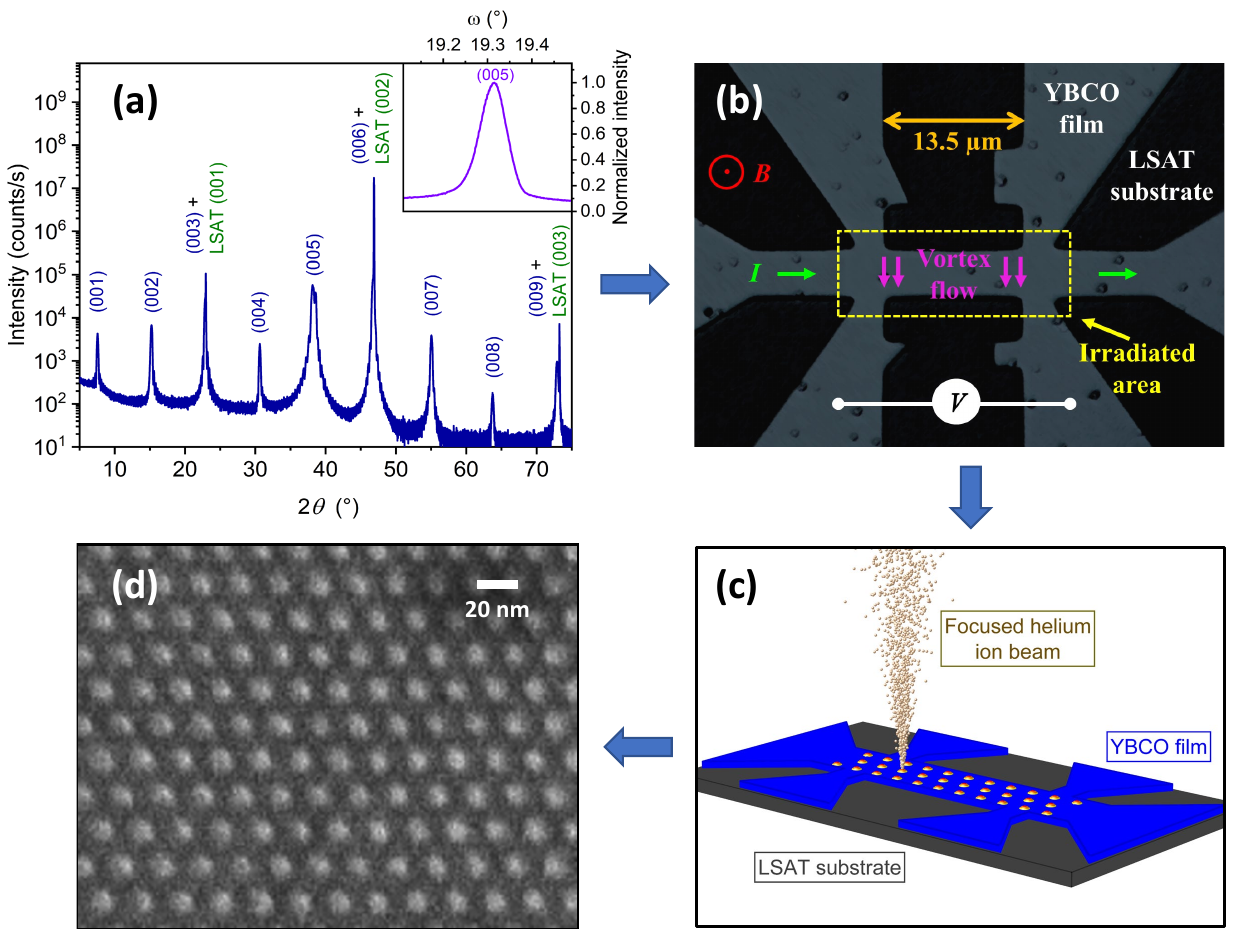}
  \end{center}
  \caption{Fabrication of nanopatterns by He-FIB: (a) X-ray diffraction $\theta$--$2\theta$ scan of the YBCO film before patterning (to create bridge H20); inset shows an $\omega$ scan of the (005) peak. (b) Optical micrograph of the bridge H20 with electrical connections after He-FIB irradiation within the area depicted by the yellow rectangle. (c) Sketch of the He-FIB patterning process. (d) Secondary electron helium ion microscope (HIM) image of a reference sample with a hexagonal $a=20\units{nm}$ array after irradiation with an excessive dose $D=60\units{ki/dot}$. The white spots are blister-forming amorphous regions.}
  \label{fig:sample}
\end{figure*}

For the studies presented here, thin YBCO films were grown epitaxially on (100)-oriented single crystal (LaAlO$_3$)$_{0.3}$(Sr$_2$AlTaO$_6$)$_{0.7}$ (LSAT) substrates by pulsed laser deposition. A 20-nm-thick Au film was deposited \emph{in situ} by electron beam evaporation right after YBCO film growth, which serves for electrical contacting. The film thicknesses (see Table~\ref{tab:samples}) are small enough to permit complete penetration of the He-FIB. This guarantees a low level of angular and energy dispersion in the columnar defect (CD) array, a necessary condition for observing matching effects \cite{NIEB01}.  An excellent $c$-axis orientation of the films is confirmed by X-ray diffraction. Figure \ref{fig:sample}(a) shows a $\theta$--$2\theta$ scan of the 37-nm-thick YBCO film, and the inset shows the rocking curve ($\omega$ scan) of the YBCO (005) peak with a full width at half maximum (FWHM) of $0.08\,^\circ$.

\begin{table*}[!htb]
  \caption{Parameters of YBCO microbridges with pinning arrays reported in this paper. The length indicates the distance between the voltage probes. The critical temperature $T_c$ is determined as the resistive transition's inflection point. The first magnetic matching field $B_1$ is calculated from Eq.~(\ref{eq:matching}) and confirmed by $I_c(B)$ and $R(B)$ measurements.}
  
  \begin{tabular}{l c c c c c c c}
    \hline\hline
    Name & Width & Length & Thickness & $a$ & $D$ & $T_c$ & $B_1$ \\
         & ($\mu$m) & ($\mu$m)& (nm) & (nm) & (ki/dot) & (K) & (T) \\
    \hline
    H30 & 8.0 & 20.0 & $26 \pm 1$ & $30 \pm 0.6$ & 10 & 75   & 2.7\\ 
    H25 & 6.5 & 17.0 & $26 \pm 1$ & $25 \pm 0.5$ & 10 & 63   & 3.8\\ 
    H20 & 5.0 & 13.5 & $37 \pm 2$ & $20 \pm 0.5$ & 10 & 52   & 6.0\\ 
    H15 & 4.0 & 10.0 & $26 \pm 1$ & $15 \pm 0.4$ & 10 & n/a  & (10.6)\\ 
    \hline\hline
  \end{tabular}
  \label{tab:samples}
\end{table*}

To fabricate microbridges on each chip, we used a combination of photolithography and Ar ion milling to pattern the Au/YBCO bilayer. Subsequently, the Au layer was removed with Lugol's iodine from the YBCO bridge structures to be exposed to He-FIB irradiation. The superconducting properties of the YBCO films did not change after patterning into microbridges. An optical image of a microbridge with an illustration of the electrical connections is shown in Fig.~\ref{fig:sample}(b). The dimensions of different bridges are chosen to maintain the same number of irradiated dots for various pinning site spacings and to minimize the irradiation time for the denser lattices (see Tab.~\ref{tab:samples}).

\subsection{He-FIB irradiation}

The He-FIB irradiation was performed in a HIM (Zeiss Orion NanoFab) at the University of T\"ubingen. The sample chips were attached to aluminum sample stubs and grounded with silver paste. After mounting in the HIM, they were thermally acclimated for roughly 60\,min before the irradiation procedure began. For simplicity, we will refer to the number of ions hitting a certain spot on the surface of the sample as the ``dose'' $D$, which will be given in units of 1000 ions per dot (ki/dot). In Fig.~\ref{fig:sample}(c), the fabrication of a hexagonal array of CDs by He-FIB is illustrated.

Notably, the dose $D = 10\,$ki/dot was carefully selected to induce point defects and locally suppress $T_c$ in the YBCO film, while avoiding amorphization within the CDs. To achieve this, the dose must not exceed a critical amorphization dose $D_c\approx 13\,$ki/dot (see Sec. \ref{sec:dose}). However, if  $D<D_c$, the areas subjected to irradiation do not exhibit any visible surface degradation and cannot be seen by HIM imaging. To monitor the CD array fabrication process and visualize it, an auxiliary sample [shown in Fig.~\ref{fig:sample}(d)] was irradiated with a dose $D \gg D_c$. The scanning transmission electron microscopy (STEM) study presented in Sec. \ref{sec:dose} reveals that, for He-FIB doses above the amorphization threshold, the amorphous region causes local swelling, resulting in a bulge of the surface, as shown in Fig.~\ref{fig:sample}(d).

Apart from the number of impinging ions, the lateral intensity profile of the He ion beam is also crucial. For our typical HIM parameters [a spot control (SC) value of 5 and an aperture of $10\units{\mu m}$], we assume a Gaussian intensity profile with an estimated FWHM beam diameter $\varnothing \sim 9\,$nm. The SC value for a given aperture diameter determines the width of the beam diameter. A higher SC value produces a narrower beam profile at the expense of lower ion fluence and therefore longer irradiation time. For example, a narrow beam is required to draw lines across a YBCO microbridge to create Josephson junctions \cite{CYBA15,MULL19}. On the other hand, a lower SC value widens the beam and can lead to unwanted defects in the inter-CD areas, as discussed in section \ref{sec:spot_control}. 

Throughout this work, we used a He-FIB with an energy of $30\units{keV}$. The irradiated area is limited by the field of view (FoV) of the HIM, and for optimum resolution of the HIM's beam deflection circuitry, the FoV should be $30 \times 30\units{\mu m^2}$ or smaller. This provides a beam positioning accuracy of approximately $0.5\units{nm}$. 

Several hexagonal arrays have been produced with spacings $a=30,\,25,\,20,\,15\,$nm and doses ranging from $D = 1$ to $10\,$ki/dot. The different values of $a$ were chosen to test the ultimate resolution that can be achieved. Here, we present the results for $D = 10\units{ki/dot}$, which demonstrates the strongest pinning. 

\subsection{Transport measurements}

The electrical measurements were carried out in a physical properties measurement system equipped with a 9-T superconducting solenoid and a variable temperature insert with a Cernox resistor for in-field temperature control (Quantum Design). For all experiments, the magnetic field was oriented perpendicular to the sample surface ($x,y$-plane), that is, the field is along the $z$ axis. To rule out thermoelectric signals, resistance measurements were taken with constant excitation currents of 1.2--2.4\,$\mu$A in both polarities. The magnetic field was swept in both polarities at fixed temperatures, and perfect mirror symmetry was observed. Furthermore, the results are independent of the magnetic field sweep direction and no hysteresis \cite{ZECH17a} was observed.

The critical current $I_c$ is determined from isothermal voltage $V$ versus current $I$ measurements using a voltage criterion of $135\units{nV}$, which is above the noise floor of $40\units{nV}$ and corresponds to an electric field $E = 100\units{\mu V/cm}$. Supplementary resistance measurements in magnetic fields up to 12.5\,T were performed in an Oxford Instruments helium bath cryostat.

\section{Simulations}

\begin{figure*}[!tb]
  \begin{center}
    \includegraphics[width=\textwidth]{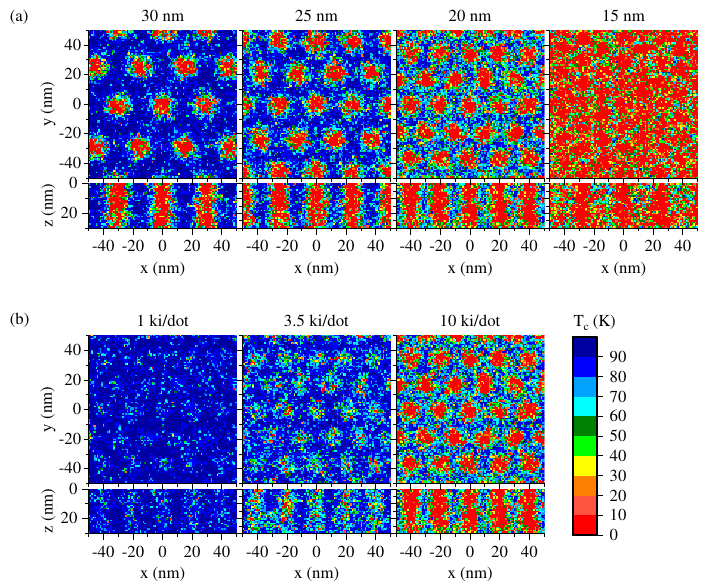}
  \end{center}
  \caption{Monte Carlo simulations of $T_c$ suppression induced by He-FIB irradiation (with $\varnothing=9\,$nm) of YBCO; only the central parts of larger hexagonal patterns are shown. The $x$-$y$ plots show cuts at $z = 15$\,nm depth below the film surface, while the $x$-$z$ plots show cuts at $y = 0$\,nm.  (a) Columnar defects (CD) lattices with various lattice spacings; $D=10\,$ki/dot. (b) CD lattices with $a = 20$\,nm and various $D$.}
  \label{fig:simus}
\end{figure*}

Our experiments were preceded by simulations of the expected $T_c(x,y,z)$ suppression patterns, which are summarized in Fig.~\ref{fig:simus}. Briefly, atom displacement cascades due to He-FIB irradiation were calculated using the SRIM/TRIM program \cite{ZIEG10}. Individual He$^+$ atom's impact points were shifted to simulate the Gaussian beam intensity distribution and the hexagonal irradiation pattern. The resulting three-dimensional landscapes of defects can be converted to local $T_c$ values by using experimental data from uniformly large-area-irradiated thin YBCO films as calibration \cite{MLET19}. The size of the simulation cells is $1 \times 1 \times 1$\,nm$^3$, similar to the in-plane and larger than the out-of-plane coherence lengths of YBCO. 

In Fig.~\ref{fig:simus}(a), the red dots indicate hexagonal arrays of well-formed nanocolumns with suppressed $T_c$. They are well separated down to $a = 20$\,nm, leaving a superconducting matrix in between. Because of the straggling of the ion trajectories, these inter-CD regions have a finite but smaller defect density and hence also slightly reduced $T_c$ values. In contrast, the simulation of an array with $a=15$\,nm reveals little $T_c$ contrast between defect columns and inter-CD regions, making the observation of matching effects unlikely.

Furthermore, proper dose selection is crucial, as demonstrated in Fig.~\ref{fig:simus}(b) for a lattice with spacings $a=20$\,nm. Doses of $D\ll 10\units{ki/dot}$ do not entirely suppress superconductivity. They only decrease $T_c$ locally, resulting in a weaker pinning potential, as shown in Fig.~\ref{fig:dose_SC}(a). Conversely, doses $D > 10\units{ki/dot}$ provide no additional benefit since the suppression of $T_c$ is already maximal. On the contrary, an excessive number of defects causes crystal lattice instabilities, which spread into neighboring regions and widen the defect columns. 

\section{Results and discussion}

\subsection{Optimization of irradiation parameters}

\subsubsection{Irradiation dose}
\label{sec:dose}

\begin{figure*}
   \begin{center}
     \includegraphics[width=\textwidth]{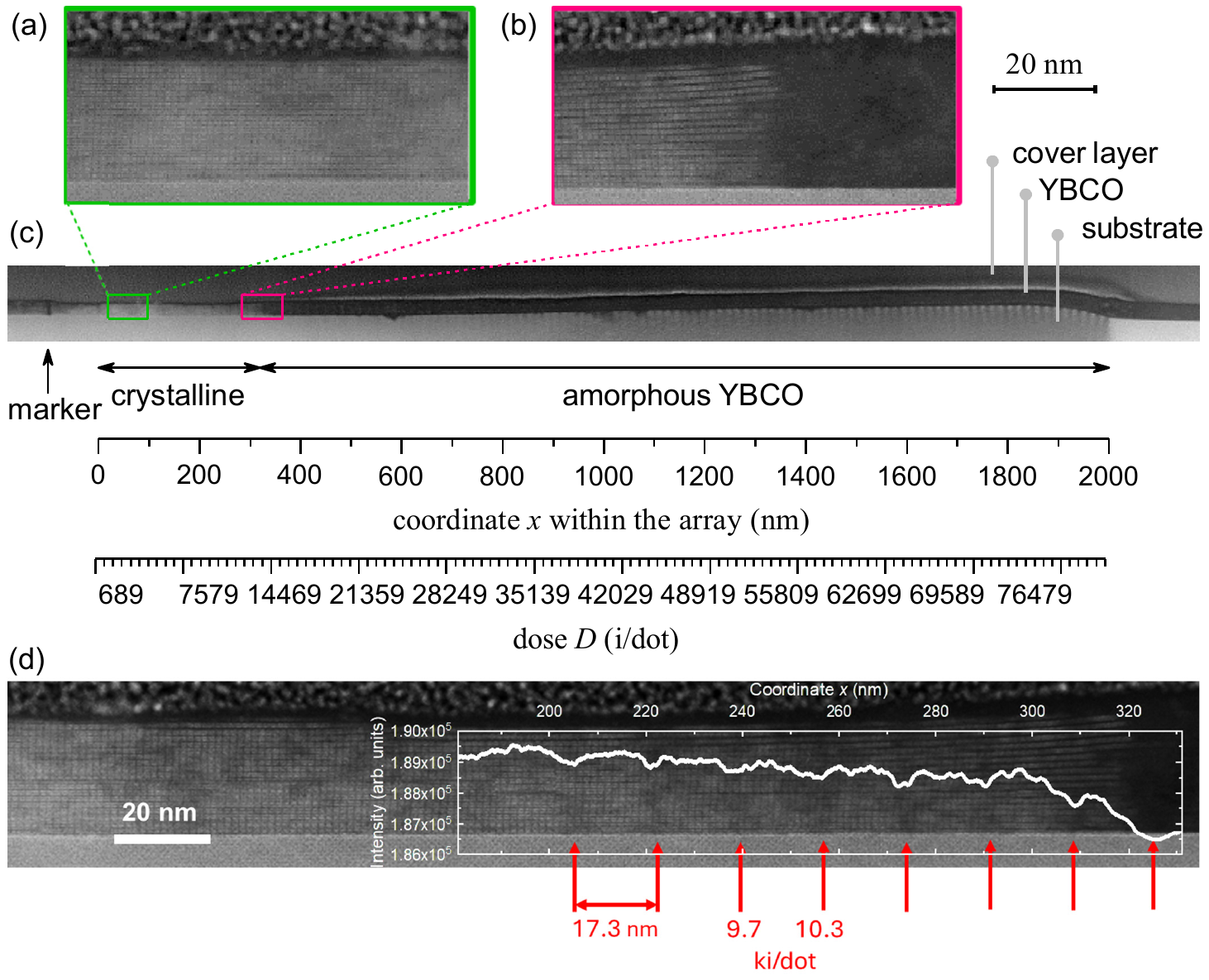}
   \end{center}
   \caption{%
   Cross-sectional High-angle annular dark field STEM image of the lamella extracted from a hexagonal array of CDs with the lattice constant $a=20\units{nm}$, length $2000 \units{nm}$ in the $x$-direction. The array has a dose gradient along the $x$ axis given by $D(x) = 689\units{i/dot} \cdot (N+1) = 689\units{i/dot}\cdot(1+x/a')$, where $N=0, \ldots,115$ is the CD column number in the gradient direction and $a'=\sqrt{3}/2a\approx17.32\units{nm}$ is the distance between the CDs in the $x$ ($D$ gradient) direction. (a),(b) Zoomed images in the region with preserved crystal structure and in the region with onset of amorphization ($x_c\approx320\units{nm}$, $D_c\approx 13\units{ki/dot}$).  (c) Overall view of the entire array. The scales below the image show the coordinate $x$ within the array and the corresponding dose $D(x)$. Each small tick on the $D$ axis corresponds to a position of a defect column (i.e., the distance between the ticks is $a'$). (d) The intensity profile (white) overlaid onto a selected part of the lamella's STEM image shows a slight contrast modulation with the periodicity $a'$. The red arrows indicate the locations of the irradiated spots.}
   \label{fig:TEM}
\end{figure*}

To maintain the crystalline framework of YBCO and still suppress its superconductivity, it is essential to determine the optimal number of 30 keV He$^+$ ions (referred to as dose $D$) to irradiate one dot of the pinning lattice. If the dose exceeds the threshold value $D_c$, it causes the breakdown of the crystalline structure, making the YBCO film amorphous. To identify $D_c$, a hexagonal array of dots was used with a distance of $a=20$\,nm. The dots were irradiated with increasing doses in steps of 689 i/dot from row to row, linearly from 0.689 ki/dot to 80 ki/dot, leaving the other irradiation parameters the same as for sample H20.

This array was characterized by aberration-corrected STEM in a probe-corrected FEI Titan 60-300 operated at 300 kV, equipped with a CEOS corrector for the condenser system, which provides a probe size below 1 \AA. A high-angle annular dark field in STEM was used to provide Z-contrast images. The cross-sectional specimen was prepared by Ga$^+$-FIB in an FEI Helios 650 Dual Beam FIB system. The bright bubble-like contrast on top of the YBCO film in Figs.~\ref{fig:TEM}(a) and (b) is the Pt-C deposit grown by focused electron beam induced deposition to protect the YBCO film during transmission electron microscopy (TEM) lamella preparation in SEM-FIB equipment. This is a standard procedure with negligible impact on the film quality due to the low dose and low energy of the electron beam used. The same type of coverage is used in the highly crystalline pristine areas, which does not evidence any apparent amorphization or beam damage.

The cross-sectional STEM image of the entire irradiated region is shown in Fig.~\ref{fig:TEM}(c). At low doses, the crystalline structure of YBCO is preserved, as can be seen in the magnified detail depicted in Fig.~\ref{fig:TEM}(a). At a critical dose, $D_c\approx 13\units{ki/dot}$, the crystalline structure of YBCO disappears in the STEM image, indicating the amorphization of the film, as displayed in Fig.~\ref{fig:TEM}(b). Note that at $D \lesssim D_c$, the YBCO layer already begins to swell, but the surface of the substrate remains flat. Only at substantially higher $D$ is the substrate also damaged, and individual beam tracks become clearly visible on the right in Fig.~\ref{fig:TEM}(c). This observation proves that the reported vortex-matching effects are not caused by local strain from the substrate, but rather originate from subtle modifications of the YBCO structure. 

The intensity profile for $D \lesssim D_c$ shown as an overlay to the STEM image in Fig.~\ref{fig:TEM}(d) reveals a reduction with increasing $x$ (i.e., increasing $D$) that begins when the YBCO lattice expands and reaches its lowest value in the amorphized region. Intriguingly, it can clearly be seen that for $D < D_c$, the layered structure of YBCO is conserved, yet the intensity profile shows little dips with a periodicity of  $a'=\sqrt{3}/2a$. Such an intensity depression is probably caused by the enhanced disorder in the CD channels. The width of the dips is up to 10\,nm, supporting our previous estimate for the diameter of the CDs.     

For lower dose values that do not cause visible damage, no signatures of the irradiation are visible in STEM images. However, the CDs created with lower doses strongly influence the behavior of magnetic flux quanta, as seen in electric transport measurements. The observation of magnetic-field commensurability effects with the flux-line lattice is a direct probe of efficient vortex pinning in periodic pinning landscapes. Minima of the resistivity and maxima of the critical current appear at so-called matching fields
\begin{equation}
  B_k = k \Phi _0/A,
  \label{eq:matching}
\end{equation}
where $k$ is an integer or rational number of vortices contained within the unit cell of the hexagonal pinning array with area $A=\frac{\sqrt{3}}{2}a^2$. The field $B_1$ corresponds to the flux density of one vortex per pinning site. When the magnetic field $B > B_1$, one of two scenarios may occur: either some or all CDs are filled with multiquanta vortices when the CD radius is larger than the (in-plane) coherence length or interstitial vortices enter between the CDs \cite{BUZD93}.

\begin{figure*}[tb]
    \centering
    \includegraphics[width=\textwidth]{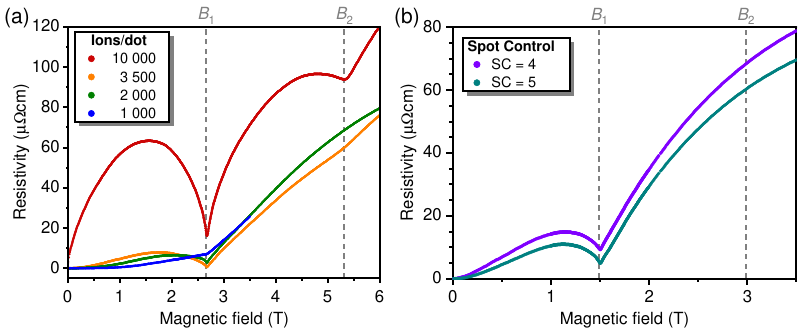}
    \caption{Vortex-matching effects in the resistivity versus field curves of irradiated YBCO bridges. (a) Magnetoresistivity of four samples at $T/T_c \sim 0.96$ after irradiation with an $a = 30$\,nm hexagonal pattern using SC\,=\,4, and doses from 1 to 10\,ki/dot. (b) Comparison of different spot control (SC) settings of the HIM. Two hexagonal pinning arrays with $a = 40$\,nm were prepared with a dose $D = 2.0$\,ki/dot on the same substrate. The magnetoresistivity at 86.5\,K ($T/T_c \sim 0.96$) is plotted for SC\,=\,4 and 5.} \label{fig:dose_SC}
    \end{figure*}

The question of how lower doses $D \ll D_c$ influence the efficiency of the pinning lattices was investigated in several microbridges prepared on the same substrate. The bridges were patterned with a hexagonal $a=30$\,nm CD lattice using different doses but otherwise identical conditions. Figure~\ref{fig:dose_SC}(a) shows the resistivity $\rho$ versus the applied magnetic field $B$ of these bridges. Because $T_c$ decreases with increasing $D$, the data were taken at the same normalized $T/T_c \sim 0.96$. Although commensurability effects are barely visible at $D = 1$\,ki/dot, they appear at higher doses. Still, the $\rho(B)$ curves of the bridges with $D \le 3.5$\,ki/dot are similar and do not show matching dips at $B_2$. Only for $D = 10$\,ki/dot, is there a clear matching effect at $B_2$, and, moreover, the minimum at $B_1$ is remarkably pronounced. These experimental results confirm the simulation results presented in Fig.~\ref{fig:simus}(b) and point to an optimal dose $D = 10$\,ki/dot for our experiments.

\subsubsection{Spot Control}
\label{sec:spot_control}

 The SC value specifies the strength of the HIM's condenser lens, thus setting the position of the crossover point in the ion optical column. For a given aperture diameter, a higher SC value thus produces a narrower beam profile at the expense of lower ion fluence and longer irradiation time.  For SC\,=\,5 used in this study we estimate a beam diameter of FWHM $\varnothing \approx 9\,$ nm on the sample surface. At a dose of 10\,ki/dot the dwell time of the beam for a single spot is about 3\,ms, and the total irradiation time of the bridge is approximately 40 minutes for the lattice spacing $a = 30$\,nm.

The processing time can be reduced by a factor of about 2 using SC\,=\,4. To test the consequences, the resistivities at 86.5\,K are compared as a function of the applied magnetic field in Fig.~\ref{fig:dose_SC}(b). The two bridges with a hexagonal CD pattern of $a = 40$\,nm were manufactured from the same YBCO film and treated identically, except for the choice of SC. Despite the similarity of the vortex-matching characteristics, the resistivity of the SC\,=\,4 bridge is higher due to a wider tail of the ion beam profile, resulting in a higher defect density between the irradiated spots. 

\subsection{Resistivity and critical current}

\begin{figure*}[tb]
  \begin{center}
    \includegraphics[width=\textwidth]{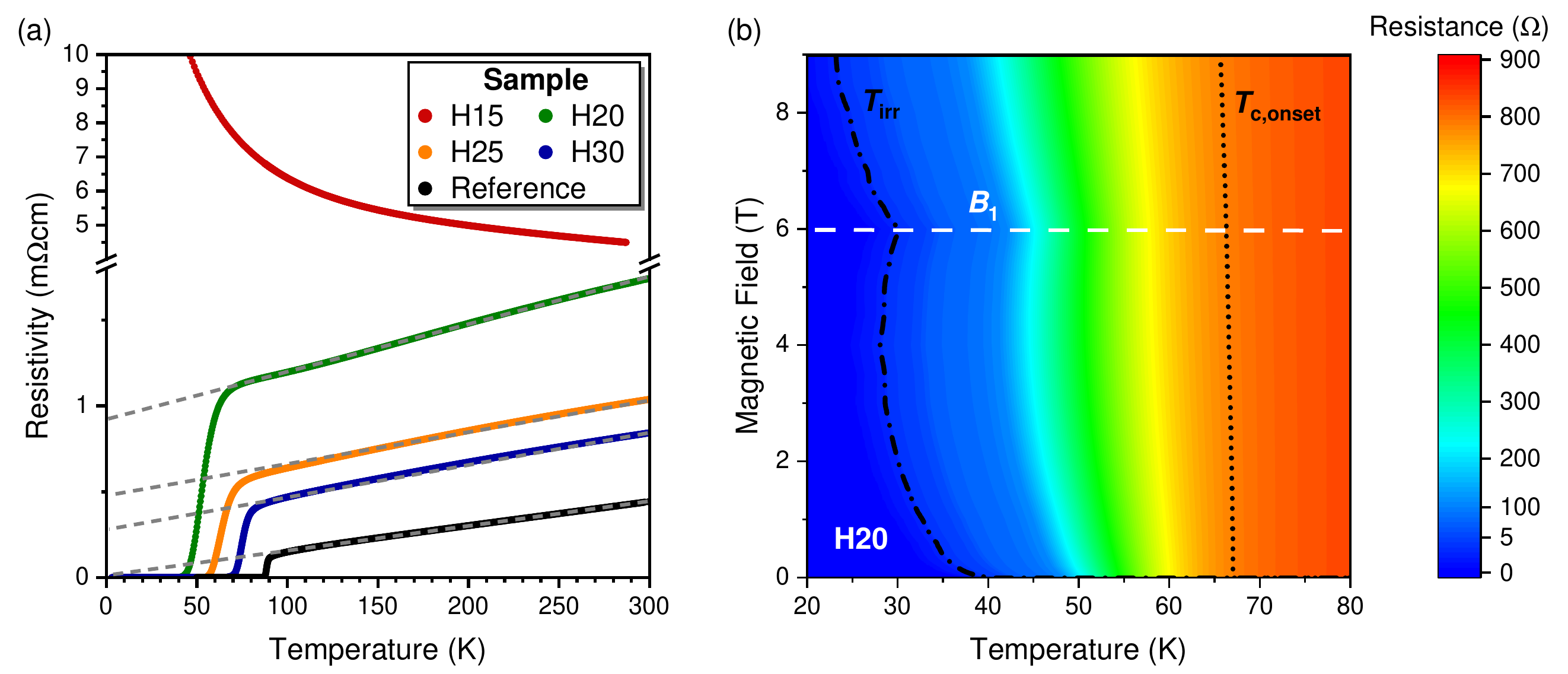}
  \end{center}
  \caption{%
    (a) Resistivity versus temperature (at $B=0$) of YBCO microbridges patterned with a hexagonal array of CDs of various spacings $a$ by He-FIB irradiation with dose $D=10\units{ki/dot}$. Extrapolations of the normal-state resistivities are shown by dashed lines. The black line shows the data of a nonirradiated reference film. (b) Color-coded contour plot of the resistance of sample H20 as a function of temperature and magnetic field. The dotted line represents the onset of superconductivity, $T_{c, onset}$, taken as 90\% of the normal-state resistance value at 100\,K, $R(100\,\text{K})$. The dashed-dotted line estimates the irreversibility line $T_{irr}$, using a criterion of 0.1\% of $R(100\,\text{K})$. 
    }
  \label{fig:R(T)}
\end{figure*}

The microbridges described in Tab.~\ref{tab:samples} were further studied by electrical transport measurements. Figure~\ref{fig:R(T)}(a) depicts the resistivity $\rho(T)$ of all four microbridges with arrays written using $D=10\units{ki/dot}$ together with a nonirradiated reference film.
The samples with $a$ = 30, 25, and 20\,nm exhibit a metallic temperature dependence in the normal state and a critical temperature that decreases with smaller spacing. An invariance of the slope indicates that no oxygen is lost during irradiation \cite{LANG10R}, while the rising offset of a straight line extrapolated from the normal state indicates an increase in the number of defects in the percolation paths between the CDs. This observation is consistent with the decline of $T_c$ and the simulation results presented in Fig.~\ref{fig:simus}(a).

In contrast, the array with $a=15\units{nm}$ displays a semiconducting $\rho(T)$ dependence and does not become superconducting even at our lowest measurement temperature of $2\units{K}$. When the CDs are placed at such a close spacing, they begin to overlap and impede the supercurrent between the CDs. This observation from the resistivity measurements for variable $a$ supports our estimate of the He-FIB's fluence profile $\varnothing = 9\units{nm}$ in the simulations.

In order to further investigate the superconducting transition of sample H20, we measured $R(T)$ for different magnetic fields, as shown in Fig.~\ref{fig:R(T)}(b). The dotted line represents the onset of superconducting percolation paths between the CDs, $T_\mathrm{c,onset}(B)$, which is defined as $R(T_\mathrm{c,onset},B) = 0.9 R(100 \units{K},B)$. The steep slope $-dB/dT_\mathrm{c,onset}$ implies that the upper critical field $B_{c2}(0\units{K})$ is very high. Remarkably, no features in $T_\mathrm{c,onset}(B)$ are observed at the first matching field $B_1=6.0\,$T (horizontal dashed line in Fig.~\ref{fig:R(T)}(b)). This suggests that in the strong fluctuation and vortex liquid regime near $T_\mathrm{c,onset}(B)$, the current paths between the CDs are basically unaffected by the pinning array and the slope of $T_{c,onset}(B)$ is comparable to that of nonirradiated films. In contrast, we observe a distinct peak of the irreversibility temperature $T_\mathrm{irr}(B)$, defined as $R(T_\mathrm{irr},B) = 10^{-3} R(100\units{K},B)$, right at $B_1$. This indicates that the onset of dissipation caused by moving vortices is significantly shifted to higher temperatures when the vortices can arrange themselves commensurably with the CD array. 

\begin{figure*}[tb]
  \begin{center}
    \includegraphics[width=\textwidth]{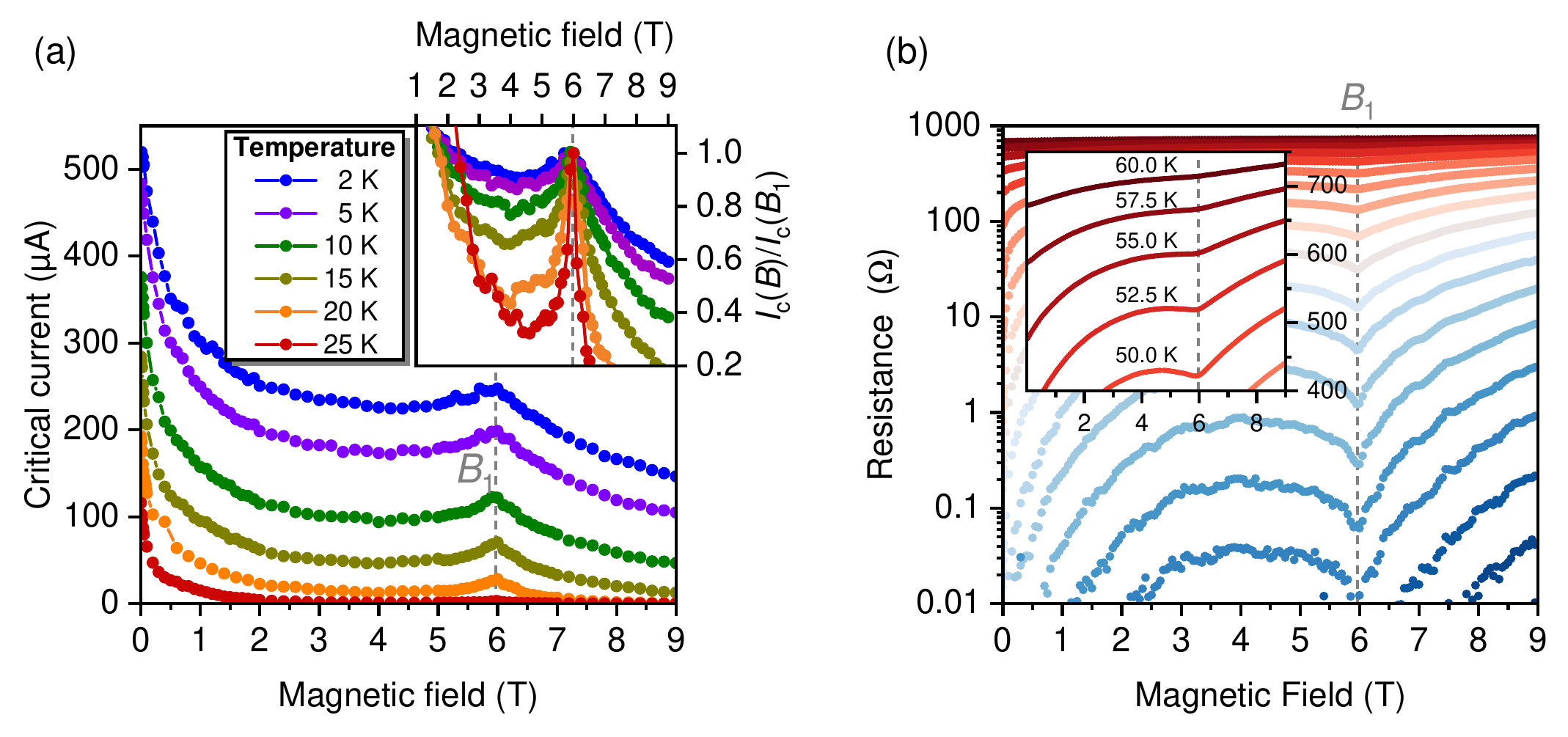}
     \end{center}
  \caption{%
    Effects of vortex commensurability in sample H20. (a) Critical current vs. applied magnetic field at various temperatures. The inset shows the critical current normalized to its value at the matching field $B_1$.  (b) Resistance vs. applied magnetic field at temperatures from 17.5\,K (dark blue symbols) to 60\,K (dark red symbols) in 2.5\,K steps. The inset shows details at higher temperatures.
  }
  \label{fig:transport}
\end{figure*}

The vortex-matching effects can be investigated more directly by critical current $I_c(B)$ (static probe) or resistance $R(B)$ (dynamic property of moving vortices) measurements. The $I_c(B)$ curves in Fig.~\ref{fig:transport}(a), taken at six different temperatures from 2 to 25\,K, show peaks at $B_1 = 6.0\,$T, which remain visible even down to $T = 2\,$K. The inset of the figure highlights these peaks by normalizing $I_c$ to $I_c(B_1)$. As the temperature increases, the pinning force of intrinsic defects decreases faster than that of CDs \cite{KHAL93}, which makes the peaks even more pronounced. It is worth noting that previous works \cite{MOSH11M,LIN96b,CAST97,AVCI10,SWIE12,CORD13,HAAG14,POCC15,SERR16,ZECH17a,DOBR18,ZECH18,AICH19,YANG19,YANG22} did not identify commensurability effects at such high magnetic fields. Furthermore, the vortex matching observed in our samples continues to occur even at very low temperatures. These observations are a natural consequence of ultradense and strong pinning sites created by He-FIB irradiation. 

From Fig.~\ref{fig:transport}(a) it is evident that the pinning effect remains strong not only at $B_1$ but also throughout the field range $0 \le B < B_1$, where commensurability effects are less significant. The minimum value of $I_c$, which roughly corresponds to the weakest pinning sites, is reached at about $4\units{T}$ at low temperatures. At the lowest temperature $T=2\,$K, $I_c(B=4\,\units{T})$ is only half the value of $I_c(B=0)$. However, for $B>B_1$ the pinning becomes less effective and $I_c$ decays more rapidly. While at $B < B_1$ every vortex can, in principle, be pinned solely in a CD, this is not the case for $B > B_1$. Then an arrangement of multiquantum vortices in the CDs or interstitial vortices occurs, which is energetically less favorable. Thus, the value of $B_1$ sets an upper limit to the typical range of strong pinning that is relevant for practical applications. This implies that shifting $B_1$ to high magnetic fields is essential, which is possible with the method presented here.

On the high-temperature side, up to about 60\,K, the vortex-matching effects are probed via reduced vortex mobility, resulting in minima of the resistance as shown in Fig.~\ref{fig:transport}(b). This suggests that even at 80\,\% of the normal-state resistance value $R(100\,\text{K})$, correlations between vortices do exist. Hence, the vortex ensemble is not entirely melted but instead moves through a plastic vortex flow \cite{REIC17R}. Voltage-current isotherms in sample H30 indicate that the interaction of artificial and intrinsic defects results in an ordered Bose glass state \cite{BACK22,AICH23}.

\begin{figure}[htb]
  \begin{center}
    \includegraphics[width=\columnwidth]{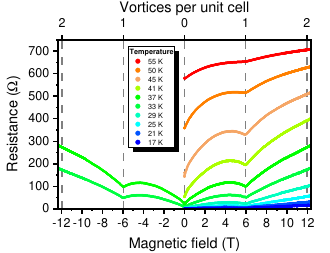}
  \end{center}
  \caption{%
    Resistance vs. applied magnetic field at various temperatures of the YBCO microbridge H20 patterned with a hexagonal array of CDs (spacing $a = 20$\,nm and dose $D=10\units{ki/dot}$). Some data were collected in both polarities of the magnetic field and confirm the symmetry $R(B) = R(-B)$. 
  }
  \label{fig:RB_full}
\end{figure}

The magnetoresistance of sample H20 was investigated up to 12.5\,T in a separate series of measurements. Figure \ref{fig:RB_full} shows the $R(B)$ curves at temperatures from 17 to 55\,K. Some measurements were taken at both magnetic field polarities to confirm the perfect mirror symmetry of the data. We do not find matching features [dips in $R(B)$] at $B_2 = 12.0$\,T, though minima in $R(B)$ at $B_2$ are visible for the H30 bridge with a wider hexagonal $a = 30$\,nm CD array (see Fig.~\ref{fig:dose_SC}(a)). In addition, $I_c(B)$, measured on the H20 bridge at lower temperatures, did not reveal any commensurability signatures at $B_2$. One possible explanation is that the dynamics of interstitial vortices changes when the channels between the CDs become narrow. Such details of vortex-matching properties in ultradense pinning landscapes deserve further investigation.

\section{Conclusions}

To summarize, irradiating thin YBCO films with a 30-keV focused helium ion beam of undercritical dose $D<D_c$ is an excellent approach for introducing user-defined patterns of suppressed critical temperature. After evaluating the appropriate irradiation conditions, we demonstrated the ability to create ultradense hexagonal pinning arrays with a spacing as low as $a=20\units{nm}$. These arrays exhibit strong pinning across a wide range of magnetic fields up to the first matching field of $6\units{T}$. The overlap of neighboring CDs sets the present experimental minimum separation of the pins at $15\units{nm} < a \leq 20\units{nm}$. However, it is possible that further adjustment of irradiation parameters could slightly reduce this constraint.

Our experiments have shown that vortex-matching effects can be observed even at low temperatures of 2 K, where superconductivity is stable and not affected by fluctuations. This paves the way for designing more effective pinning landscapes that can operate at several tesla and at temperatures much lower than the critical temperature, avoiding the undesirable vortex liquid and fluctuation regimes. 

Our findings have expanded the range of temperatures and magnetic fields that can be studied with \emph{regular} artificial pinning landscapes. In order to bring the theoretical concepts in the field of fluxonics \cite{WAMB99,GOLO15,GOLO23,MILO07} into practical use and to facilitate high operating frequencies, it is essential to confine vortices to designed features with the smallest possible size \cite{HAST03}. With our findings, it is now possible to create complex pinning patterns with narrow spacing that can be utilized in applications such as vortex-based cellular automata for fast logic with low power consumption \cite{HAST03,MILO07}. Future developments in He-FIB instrumentation, such as a multibeam technique \cite{HOFL23}, would further augment the versatility of the presented method.

\begin{acknowledgements}
    
We are grateful for insightful discussions with John Notte (ZEISS Semiconductor Manufacturing Technology) and Gregor Hlawacek (Helmholtz-Zentrum Dresden-Rossendorf). B.B. acknowledges financial support from the Vienna Doctoral School in Physics (VDSP). O.D. acknowledges financial support from the Austrian Science Fund (FWF) under grant I6079-N (FluMag). Work by O.D. was funded by the Deutsche Forschungsgemeinschaft (DFG, German Research Foundation) under Germany’s Excellence Strategy – EXC-2123 QuantumFrontiers – 390837967. C.M. acknowledges financial support from Regional Gobierno de Aragón through Project No. E13\textunderscore23R, including FEDER funding and from the EU Horizon 2020 programme under Grant Agreement No. 823717-ESTEEM3. The authors acknowledge the use of instrumentation as well as the technical advice provided by the National Facility ELECMI ICTS, node “Laboratorio de Microscopías Avanzadas (LMA)” at Universidad de Zaragoza. The research was funded in whole, or in part, by the Austrian Science Fund (FWF) Grant No. I4865-N and the German Research Foundation (DFG), Grant No. KO~1303/16-1. For the purpose of open access, the authors have applied a CC-BY public copyright license to any author accepted manuscript version arising from this submission. The research is based upon work from COST Actions CA19140 (FIT4NANO), CA21144 (SuperQuMap), and CA19108 (Hi-SCALE) supported by COST (European Cooperation in Science and Technology). 

\end{acknowledgements}

\end{document}